\documentclass[journal,twoside,web]{ieeecolor}
\usepackage{generic}
\usepackage{cite}
\usepackage{amsmath,amssymb,amsfonts}
\usepackage{algorithmic}
\usepackage{graphicx}
\usepackage{textcomp}
\usepackage{caption}

\def\BibTeX{{\rm B\kern-.05em{\sc i\kern-.025em b}\kern-.08em
    T\kern-.1667em\lower.7ex\hbox{E}\kern-.125emX}}
\usepackage{ragged2e} 
\usepackage{float}

\def\bx{\boldsymbol{x}}
\def\by{\boldsymbol{y}}
\def\bz{\boldsymbol{z}}
\def\bD{\boldsymbol{D}}
\def\bI{\boldsymbol{I}}
\def\bV{\boldsymbol{V}}
\def\bX{\boldsymbol{X}}
\def\bY{\boldsymbol{Y}}
\def\bZ{\boldsymbol{Z}}

\def\bS{\boldsymbol{S}}
\def\balpha{\boldsymbol{\alpha}}
\def\bbeta{\boldsymbol{\beta}}
\def\bgamma{\boldsymbol{\gamma}}
\def\bmu{\boldsymbol{\mu}}

\usepackage[table,xcdraw]{xcolor}
\RequirePackage[colorlinks,citecolor=blue,urlcolor=blue]{hyperref}
\definecolor{olivergray}{HTML}{CCCCCC}
\definecolor{oliverblue}{rgb}{0,0.541,0.855}
\begin{document}

\title{Personalized Longitudinal Assessment of Multiple Sclerosis Using Smartphones}

\author{Oliver Y. Ch\'en,
Florian Lipsmeier, Huy Phan, Frank Dondelinger, Andrew Creagh, \\ Christian Gossens, Michael Lindemann,  
and Maarten de Vos
\thanks{
This work was funded by F. Hoffmann-La Roche Ltd and NIHR Oxford Biomedical Research Centre (BRC). The authors thank Sven Holm and Guy Nagels for helpful comments regarding early versions of the paper.}
\thanks{O.Y. Ch\'en is with the University of Bristol, Bristol BS8 2LR, UK  (e-mail: \href{mailto:olivery.chen@bristol.ac.uk}{\textcolor{oliverblue}{olivery.chen@bristol.ac.uk}}). }
\thanks{F. Lipsmeier, C. Gossens, and M. Lindemann are with Roche Innovation Center Basel, F. Hoffmann-La Roche Ltd, Basel 4070, Switzerland.}
\thanks{H. Phan is with School of Electronic Engineering and Computer Science, QMUL, Queen Mary University of, London E1 4NS, UK, and with the Alan Turing Institute, London NW1 2DB, UK.}
\thanks{F. Dondelinger was with  Roche Innovation Center Basel, F. Hoffmann-La Roche Ltd, Basel 4070, Switzerland, and is now with the Novartis Institutes for BioMedical Research (NIBR), Novartis AG, Basel 4033, Switzerland.}
\thanks{A. Creagh is with the Institute of Biomedical Engineering, University of Oxford, Oxford OX3 7DQ, UK and with the Big Data Institute, University of Oxford, Oxford OX3 7LF, UK.}
\thanks{O.Y. Ch\'en consulted for F. Hoffmann-La Roche. F. Lipsmeier is an employee of F. Hoffmann-La Roche Ltd. F. Dondelinger was an employee of F. Hoffmann-La Roche Ltd. A. Creagh was a PhD student at the University of Oxford and acknowledges the support of F. Hoffmann-La Roche Ltd. C. Gossens is an employee and shareholder of F. Hoffmann-La Roche Ltd. M. Lindemann is a consultant for F. Hoffmann-La Roche Ltd via Inovigate. H. Phan and M. de Vos have nothing to declare.}
}
\maketitle

\begin{abstract}
Personalized longitudinal disease assessment is central to quickly diagnosing, appropriately managing, and optimally adapting the therapeutic strategy of multiple sclerosis (MS). It is also important for identifying the idiosyncratic subject-specific disease profiles. Here, we design a novel longitudinal model to map individual disease trajectories in an automated way using sensor data that may contain missing values. First, we collect digital measurements related to gait and balance, and upper extremity functions using sensor-based assessments administered on a smartphone. Next, we treat missing data via imputation. We then discover potential markers of MS by employing a generalized estimation equation. Subsequently, parameters learned from multiple training datasets are ensembled to form a simple, unified longitudinal predictive model to forecast MS over time in previously unseen people with MS. To mitigate potential underestimation for individuals with severe disease scores, the final model incorporates additional subject-specific fine-tuning using data from the first day. The results show that the proposed model is promising to achieve personalized longitudinal MS assessment; they also suggest that features related to gait and balance as well as upper extremity function, remotely collected from sensor-based assessments, may be useful digital markers for predicting MS over time. 
\end{abstract}


\section{Introduction}
\label{sec:introduction}

Multiple sclerosis (MS) is a chronic autoimmune, inflammatory, and demyelinating disease of the central nervous system \cite{Reich2018multiple}. Worldwide, approximately 2.3 million people are affected by the disease \cite{milo2010multiple, browne2014atlas}, with a pooled incidence rate of 2.1 per 100,000 persons/year across 75 reporting countries, and the mean age of diagnosis being 32 years \cite{walton2020rising}.

Progression of MS commonly leads to accumulation of impairment in one or several functional domains, including upper extremity function, gait and balance, cognition, and vision \cite{kister2013natural}. Impairment to gait and balance as well as to upper extremity function can negatively impact the quality of life and the ability to perform activities of daily living \cite{yozbatiran2006motor, poole2010dexterity, lam2021real, bisio2017kinematics, zwibel2009contribution}. Previous reports suggest that up to 75–90\% of people with MS (PwMS) experience impaired gait, and up to 60–76\% of PwMS show signs of impaired upper extremity function \cite{kister2013natural, hemmett2004drives, johansson2007high, bertoni2015unilateral}.

Regular assessment of functional ability can help to guide early treatment decisions and thereby improve treatment outcomes. At present, individuals at risk of developing MS are examined based on a combination of in-clinic assessments, including checking for MS-related signs and symptoms, neuroimaging studies, and laboratory testing \cite{tsang2011multiple}. The Expanded Disability Status Scale (EDSS), rated by clinicians, measures overall disease impairment (\textit{i.e.}, disability in functional systems and ambulation) \cite{kurtzke1983rating}. Alternatively, the Multiple Sclerosis Impact Scale (MSIS-29), self-assessed by patients, captures the physical and psychological impact of the disease \cite{riazi2002multiple, hobart2001multiple, mcguigan2004multiple}. The EDSS score ranges from 0 to 10 (representing normal to death due to MS). The MSIS-29 total score (physical score plus psychological score) ranges from 25 to 145 (representing best to worse physical and psychological functions).

To frequently assess, appropriately manage, and optimally adapt the therapeutic strategy of MS, it is critical to build suitable longitudinal models. Nevertheless, there are a few challenges when developing longitudinal models for MS studies. 

First, while regular assessment of functional ability is recommended to support optimal adaptation of the therapeutic strategy \cite{rae2015quality}, clinical measures of functional ability are at present, unfortunately, only infrequently administered. This makes it difficult to align disease measurements timely or accurately with therapeutic strategy recommendations. This difficulty is further complicated by the fluctuating nature of MS symptoms \cite{rae2015quality, steelman2015infection, mills2017emerging}. Finally, the data collected from such (infrequent) evaluations are not suitable for longitudinal model development. Recently, smartphone sensor-based assessments are beginning to allow remote and frequent assessment at home without supervision. Research has shown that the remote, frequent (daily), and objective assessment of MS-related functional impairment is feasible with smartphone sensor-based assessments \cite{montalban2021smartphone, chen2021personalized}. These assessments can be typically taken at home without supervision and with minimal burden on the patient \cite{midaglia2019adherence}. The relatively low cost and high penetration rate of smartphone devices, as compared to the availability of hospital devices and physicians, make such assessments available to a large population \cite{poushter2016smartphone, bhavnani2016mobile}. Taken together, more frequent assessments of functional ability using sensor-based smartphone technology are likely to provide notable value in tracking MS-related impairment. 
	
In parallel, although recent advancements in sensor technology allow for more frequent sensor-based disease assessments, a suitable longitudinal model must address two main challenges. On the one hand, longitudinal sensor data oftentimes contain missing values. If not treated, one cannot make full use of the data and, particularly, cannot make disease assessments on days with incomplete data. On the other hand, although PwMS exhibit common patterns at the population level, personal disease prediction on novel PwMS using results discovered from others may not capture the subject-specific information of the novel PwMS. It is thus important to develop a longitudinal method that balances extrapolation and personalization.

Here, we propose here a personalized longitudinal framework to automatically assess MS over time. Through data analyses and out-of-sample testing using smartphone sensor-based data from the study ``Monitoring of Multiple Sclerosis Participants with the Use of Digital Technology (Smartphones and Smartwatches) - A feasibility Study'' \cite{montalban2021smartphone}, we demonstrate that the framework has the potential (1) to extract MS-specific digital clinical markers and (2) to assess individual MS trajectories longitudinally.

\section{Methods}

\begin{figure*}[h]
\centering
\captionsetup{justification=centering}
\includegraphics[width=180mm]{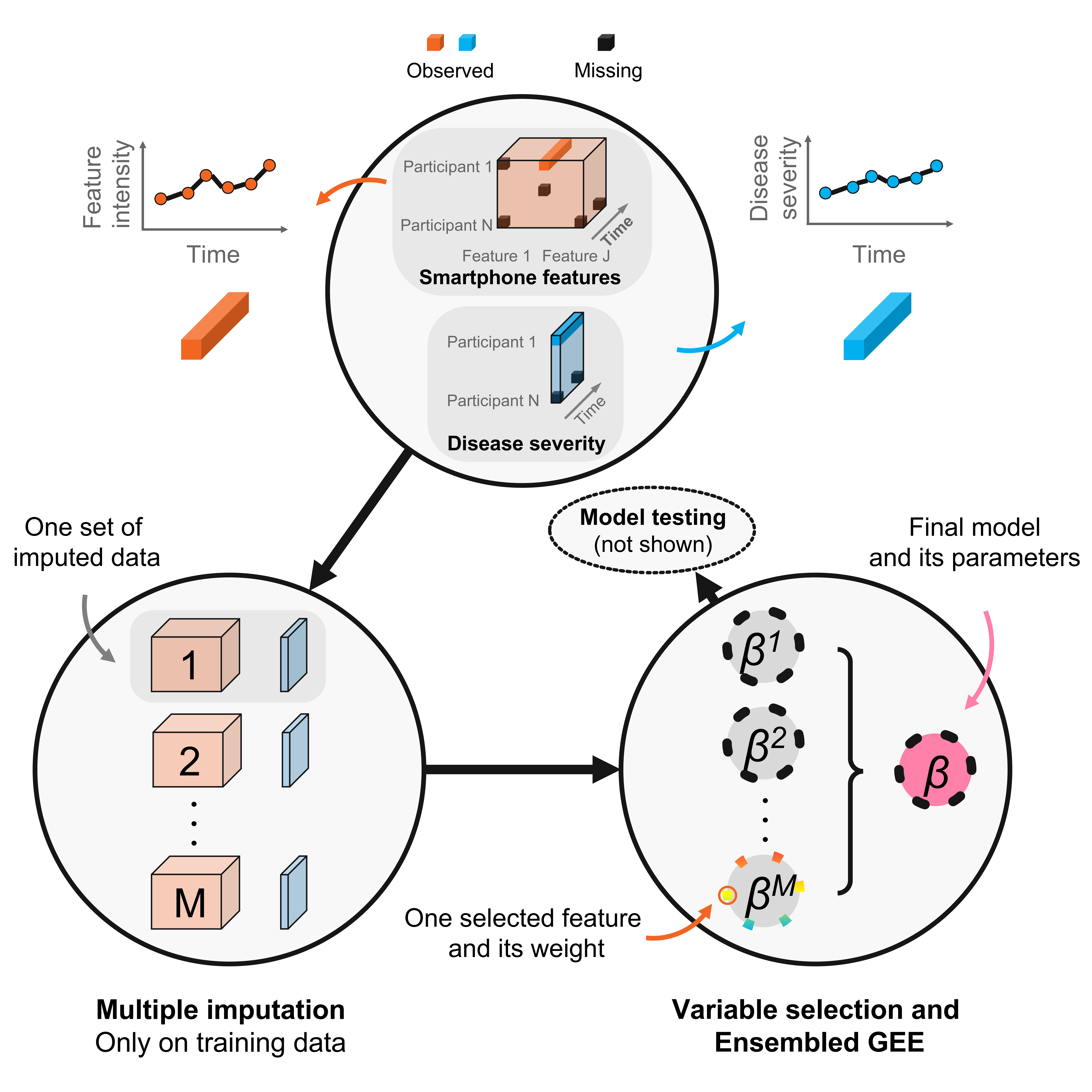}
\noindent\colorbox{olivergray}{
\parbox[c]{17.8cm} {
Top: Data organization. The smartphone data are organized in an orange cubic, where multi-dimensional feature data are collected from each participant over time. Specifically, the orange bar indicates one feature measured over time for one subject. Disease outcomes are similarly prepared, where the blue bar shows the disease scores measured over time. The black boxes refer to missing data. Lower left: Data imputation. Missing data are imputed via multiple imputation (MI). To accommodate the uncertainties in the imputed data, we compute multiple ($M$) sets of complete data. Lower right: Model development. A generalized estimation equation is fit on each of the fully imputed datasets to obtain parameter estimation. The $M$ sets of estimated parameters are then pooled to generate a unified model, which will be used for the model test. The (pooled) parameters are applied to features from the test sample to estimate outcomes. To prevent data imputation from influencing the test results, only complete observations will be used during the test (namely, MI will not be used during the test). The estimated results are then compared with the observed outcomes to verify the performance as well as reproducibility of the model.
}
}
\caption{A schematic representation of the longitudinal model.}
\smallskip
\label{Fig_1}
\end{figure*}

The structure of this section is organized as follows. In \textbf{Section} \ref{sec:2.1}, we will introduce the smartphone-based MS data used in this paper. In \textbf{Section} \ref{sec:2.2}, we will discuss how we deal with missing data in longitudinal studies. In \textbf{Sections} \ref{sec:2.3}-\ref{sec:2.4}, we introduce the model and how to make inferences about the parameters. \textbf{Section} \ref{sec:2.5} presents how to ensemble models to make out-of-sample predictions. In \textbf{Section} \ref{sec:2.6}, we introduce personalized fine-tuning to improve longitudinal disease assessment. Finally, in \textbf{Section} \ref{sec:2.7} we investigate the optimal number of imputed datasets. The data organization, imputation, model development, and parameter ensemble are summarized in \textbf{Fig.} \ref{Fig_1}.

\subsection{MS data collected remotely at home using smartphones} \label{sec:2.1}

We used data from a 24-week, prospective study (clinicaltrials.gov identifier: NCT02952911) aimed to assess the feasibility of remotely monitoring PwMS with sensor-based assessments \cite{montalban2021smartphone, midaglia2019adherence}. The current study consisted of smartphone sensor-based data collected from 76 MS patients (18–55 years; EDSS score 0–5.5). Smartphone data were collected using a preconfigured smartphone (Samsung Galaxy S7) that prompted the participants to perform daily assessments of upper extremity function, gait and balance, and cognition \cite{montalban2021smartphone, midaglia2019adherence}. The local ethics committees approved the study. Written informed consent was obtained from all participants; see \cite{montalban2021smartphone} for the exclusion criteria for MS patients and other study details.

Walking ability and upper extremity function are indispensable to support independence in activities of daily living, and their functions are often impaired in PwMS \cite{kieseier2012assessing, kierkegaard2012relationship}. Previous studies have shown that digital features related to walking ability and upper extremity functions are useful to assess PwMS \cite{creagh2020asmartphone, creagh2020bsmartphone, bourke2020gait}. Identifying and isolating the subset(s) of features that capture impairment in these functional domains and that are potentially predictive of MS disease severity, therefore, may help to improve one's understanding of the pathological aspects of MS that are related to gait and balance as well as to upper extremity function. Additionally, the selected features may inform clinical decisions that aim at preserving the functional abilities in these two domains. In parallel, the magnitude of impairment in these features may be associated with the severity of MS. For example, worsening walking ability is associated with MS disease progression and/or one or more relapses \cite{scalfari2010natural}. Similarly, worsening upper extremity function is frequently reported in PwMS and is increasingly present as the disease progresses \cite{feys2017nine}. Together, these lines of evidence suggest that gait and upper extremity function features may be potentially useful markers to longitudinally assess the severity and progression of the disease. 

The data considered in this study are objective measurements obtained from the Two-Minute Walk Test (2MWT), which assesses gait and dynamic balance, and the Draw a Shape (DaS) Test, which assesses upper extremity function. The 2MWT instructs study participants to walk as fast as possible, but safely, on even ground for two minutes without performing U-turns. The DaS Test involves drawing six increasingly complex shapes (two diagonal lines, a circle, a square, a figure-of-8, and a spiral). The reason to consider these data is twofold. 

The study also included the clinician-rated Expanded Disability Status Scale (EDSS) scores and the self-reported MS Impact Scale (29-item scale) (MSIS-29) scores. For the present study, we mainly focused on predicting MSIS-29 scores because they were collected from the same smartphone application every two weeks, and the data were available more frequently; consequently, they were more suitable for developing and evaluating a longitudinal model. Nevertheless, for completion purposes, we also evaluated and briefly discussed the model performance for predicting EDSS scores in this paper. One should, however, note that there were only a very small number of EDSS scores per subject for training and test.

\subsection{Treating missing data in longitudinal studies} \label{sec:2.2}

Missing data are common to data collection and pose challenges for longitudinal disease prediction \cite{prince2018multi, creagh2022longitudinal}. When there are missing inputs (in the design matrix) or missing outputs for the training data, model development becomes difficult. Therefore, a practical longitudinal disease predictive model needs to first treat the problem of missing data.

In this paper, we used the multiple imputation (MI) \cite{graham2009missing} to handle missing data. We adopted it primarily because it was relatively easy to control for model efficiency (see \textbf{Eq.} \eqref{eq:eq4}) by adjusting the number of imputations performed \cite{sterne2009multiple, von2016new} (see below). Additionally, it provided a principled way to estimate the uncertainty associated with the imputation \cite{rubin2004multiple, rubin1996multiple}. Finally, for practitioners, MI may be easier to implement as it is readily available in standard software. One can refer to \cite{rubin1976inference, allison2001missing, enders2010applied, little2019statistical} for additional treatments of missing data. Note that while the use of MI for randomized clinical trials warrants some caution \cite{jakobsen2017and}, the design of this study did not include any randomization into treatment groups, and hence the application of MI was straightforward; we, however, do assume that data are missing at random (MAR).

Here, we briefly outline the steps of MI. First, a column that contains missing values is regarded as a target column (and treated as a response variable), and the remaining columns of the data set are used as predictors. The missing values in the targeted column are then filled using the predicted mean matching (PMM) \cite{little2019statistical, andridge2010review}. For predictors that are incomplete in themselves, the most recently generated values are used so that the predictors are complete before making imputations for the target column. Second, the first step is repeated for every column that contains missing data. Finally, the first two steps are repeated multiple ($M$) times to obtain $M$ sets of complete data. These steps can be done using a publicly available \textit{R} package \textit{mice} \cite{van2011mice}.

\subsection{Model development using the GEE} \label{sec:2.3}

The longitudinal predictive model ensembles multiple generalized estimation equations (GEEs) followed by a subjective-specific fine-tuning (see \textbf{Fig.} \ref{Fig_1}). The choice of choosing the GEE is twofold. First, it can take both correlated and uncorrelated repeated longitudinal measurements (both within and between PwMS). Second, even if one mis-specified the correlation structure, the parameter estimates would still be consistent \cite{liang1986longitudinal}. The subject-specific fine-tuning is added to treat potential overfitting during training and underestimating for novel PwMS with severe scores during the test.

Before introducing the subject-specific aspect of the model (see \textbf{Section} \ref{sec:2.6}), let's first outline the general longitudinal predictive model. Consider $N$ PwMS, where the $i^{\text{th}}$ subject has $p_w$ gait features and $p_d$ hand function features, measured at time $t=(1,2,\ldots,T_i)$ (the subscripts $w$ and $d$ indicate walking (gait) and dexterity (hand) functions, respectively). The feature matrix for the $i^{\text{th}}$ subject is thus $\bx_i^\mathsf{T}=(\bx_{iw}^\mathsf{T}, \bx_{id}^\mathsf{T} )$, $\bx_{ij}^\mathsf{T}=(\bx_{ij1}, \ldots, \bx_{ij{p_j}} ) $), for $ j \in \{w,d \}$, and $\bx_{ijk}$ contains longitudinal measurements $\bx_{ijk}= \left( x_{ijk} (1), \ldots, x_{ijk} (T_i )\right)^\mathsf{T}$, for $1 \leq k \leq p_j$, where $\mathsf{T}$ denotes the transpose operation. Let $\by_i=\left( y_{i1}, \ldots, y_{i {T_i}} \right)^\mathsf{T}$ be the longitudinal outcome for the $i^{\text{th}}$ subject and $y_{it}$ be the outcome for the $i^{\text{th}}$ subject at time $t$.

Consider $p_w=2$ gait features from a Two-Minute Walk (TMW) test, and $p_d=51$ hand function features from a Draw a Shape (DaS) test. For each individual $i$, the number of repeated measures, or $T_i$, ranges from $1$ (two PwMS) to $23$ with a mean of $8.5$. The model also includes covariates such as sex, age, race, and site (from where data were collected). Unless explicitly stated, the features for model development discussed in this section were from $57$ randomly selected training PwMS. Features from the remaining $23$ PwMS were held out for the test. To avoid a chance split during training and test, we will perform bootstrap experiments in \textbf{Section} \ref{sec:3.1}.

We considered feature selection in two steps. The first step selected features in the training set that were significantly associated with the outcome (see \cite{chen2020building}). The second step selected features using the GEE (see \textbf{Fig.} \ref{Fig_4}).

Consider a feature $\bx$. Denote $\rho_{(\bx, \by)}$ as the result from a statistical test during the step-one feature selection between the feature $\bx$ and the outcome $\by$. For disease severity outcomes, the value of $\rho$ is the correlation between the feature and the outcome; equivalently, it could be the corresponding $p$-values. Let $\epsilon$ be a pre-specified threshold for $\rho$. Although there is a one-to-one mapping between a statistic and its $p$-value, sometimes it may be convenient to evaluate the $p$-value, sometimes it may be convenient to evaluate the correlation, and other times, one may be interested in constraining the number of features being selected. For example, in this study, we set $\epsilon$ to control the total number of selected features. 

\begin{figure}
\includegraphics[width=90mm]{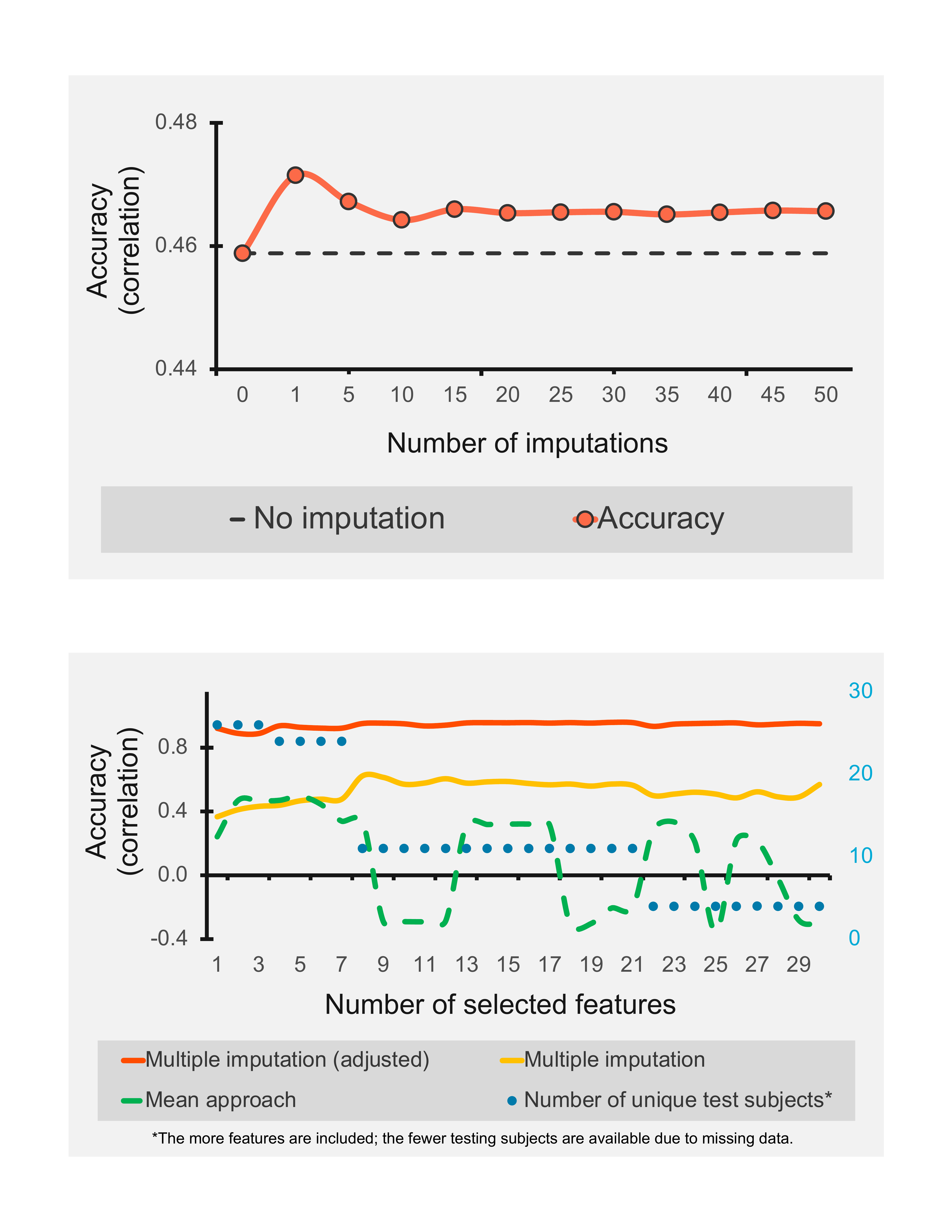}
\smallskip
\noindent\colorbox{olivergray}{
\parbox[c]{8.3cm} {Top: Model performance using different numbers of imputations. Model performance improved with imputations compared to without and generally stabilized when there were more than $15$ imputations. Bottom: Model performances for the mean approach, raw generalized estimation equations (GEE) approach, and adjusted GEE approach across different numbers of selected features. Due to missing values in features, the size of the available independent test sample (indicated by the height of the blue dots) decreased when more features were used. This is because we only considered observed features for testing (namely we did not perform MI for the test sample to prevent it from potentially biasing prediction performance). In general, the prediction accuracy using MI and GEE (indicated by the yellow line) was relatively high and stable across various numbers of features selected. The fine-tuning further improved prediction performance (see the orange line). 
}
}
\caption{Model comparison for longitudinal MSIS-29 prediction using different sets of imputations and various numbers of features.}
\label{Fig_2}
\end{figure}

Six features (including the best four smartphone features plus age and sex) were allowed in the final model; we explored other thresholds that allowed different numbers of features in the final model as a complementary analysis (see \textbf{Fig.} \ref{Fig_2}). The reason that we constrained the total number of features in the model was due to missing data in the test dataset. Although one can deal with missing data in the training set via imputation, to avoid bias brought into the test data by imputation, we chose to not perform MI on the test set and evaluated the model performance only on available observations. Naturally, if one included a larger number of features, there would be fewer numbers of PwMS left with complete features. We found that a total of six features would balance the number of features for training and the number of available test data (see \textbf{Fig.} \ref{Fig_2}). 

Formally, the (general) marginal model that studies the relationship between the features and outcomes is:

\begin{equation} \label{eq:eq1}
\mathbb{E} \left( y_{it} \vert  \bx_{it}, \bz_i  \right) 
= 
g^{-1} ( \mu + \bx_{it}^\mathsf{T}  \bS \balpha + \bz_i^\mathsf{T} \bgamma)          
\end{equation}

\medskip

\noindent
where $g(\cdot)$ is a link function,  $\mu$ is the intercept, $\bx_{it}^\mathsf{T} = [\bx_{iwt}^\mathsf{T}, \bx_{idt}^\mathsf{T}]$ and $\bx_{ijt}^\mathsf{T}= ( x_{ij1} (t), x_{ij2} (t)$, $\ldots, x_{ij {p_j}}  (t) ) $, for $j  \in \{w,d \}$. Here, $\mathcal{\bS}=\text{blockdiag} \{ \bI_w, \bI_d \}$, and $\bI_j = \text{diag} \{ i_{j1}, i_{j2}, \ldots, i_{j {p_j}} \}$, wherein $i_{jk}=1$ if $\rho_{ \left ( \bx_{jk}, \by, N-2 \right ) }   <  \epsilon$ and $0$ otherwise, where $\bx_{jk}= \left (\bx_{1jk}^\mathsf{T}, \bx_{2jk}^\mathsf{T}, \ldots, \bx_{Njk}^\mathsf{T} \right) ^\mathsf{T} $ denotes a particular feature across all PwMS over time; $\bz_i$ is a vector containing all covariates for the $i^{\text{th}}$ subject, and $\bgamma$ is its coefficient. Finally, write  $\bbeta=(\balpha, \bgamma)$.

Denote the selected features as $\tilde{\bX} = \bX \bS$ and the outcome disease scores as $\bY$. Via MI, we obtained $M$ sets of complete data $\left( \bY^{(1)}, \tilde{\bX}^{(1)}, \bZ \right), \ldots, \left( \bY^{(M)}, \tilde{\bX}^{(M)}, \bZ \right)$, where the superscript ($m$) denotes the $m^{\text{th}}$ set of imputed data.

The problem in \textbf{Eq.} \eqref{eq:eq1} can be solved using the GEE. Specifically, the estimation of $\bbeta$ can be done by solving the following score function:

\begin{equation*}
\bD_i^\mathsf{T} \bV_i^{-1} (\by_i - \bmu_i) = 0
\end{equation*}
\noindent
where $\bD_i = \dfrac{\partial \bmu_i}{ \partial \bbeta^\mathsf{T} }$, $\bmu_i$ is the expectation of $\by_i$, and $\bV_i$ is the estimate (see below) of the variance-covariance matrix of $\by_i$. Let:

\begin{equation*}
\bV_i = \Delta_i^{1/2}  \mathcal{R}_i (\alpha) \Delta_i^{1/2}
\end{equation*}

\medskip

\noindent
where $\Delta_i = \text{diag} \{ \text{var}(y_{i1} ), \text{var}(y_{i2} ), \ldots,  \text{var}(y_{i{T_i }}) \}$, $\mathcal{R}_i (\alpha)$ is a $T_i \times T_i$ “working” correlation matrix \cite{liang1986longitudinal}, and $\alpha$ represents a vector of parameters associated with a specific model for correlation of $\by_i$. When $\mathcal{R}_i (\alpha)$ is the true correlation matrix for $\by_i$, then $\bV_i =  \text{cov}( \by_i)$. When the variance-covariance structure is mis-specified, the parameter estimates would still be consistent \cite{liang1986longitudinal}.

\subsection{Inference} \label{sec:2.4}

A GEE model was fit on selected features from each imputed data set. Let $\hat{\bbeta}^m=(\hat{\beta}_1^m, \hat{\beta}_2^m, \ldots, \hat{\beta}_p^m)$ denote the estimated parameter corresponding to the $m^{\text{th}}$ set of imputed data, where $p$ (\textit{e.g.}, $p=6$) is the number of selected smartphone features (\textit{e.g.}, $q=4$) plus the number of covariates ($p-q=2$), and $1 \leq m \leq M$. Then, the $(1- \alpha)\times 100 \%$ confidence interval (CI) for the estimated $\hat{\beta}^j$ corresponding to the $j^{\text{th}}$ feature, where $1 \leq j \leq p$, is:

\begin{equation*}
\left [
\bar{\hat{\beta}}_j - t_{ \left (n-1, \dfrac{\alpha}{2} \right ) } \dfrac{s_j}{\sqrt{M}},
\hspace{5mm}
\bar{\hat{\beta}}_j + t_{ \left ( n-1, \dfrac{\alpha}{2} \right ) } \dfrac{s_j}{\sqrt{M}}
\right]
\end{equation*}

\medskip

\noindent
where $\bar{\hat{\beta}}_j = \sum_{m=1}^M \hat{\beta}_j^m / M$ and $s_j= \sqrt{\sum_{m=1}^M (\hat{\beta}_j^m - \bar{\hat{\beta}}_j)^2 /M}$.

\subsection{Model ensemble and out-of-sample prediction} \label{sec:2.5}

We obtained the ensemble parameter $\bar{\bbeta}$ from $M$ GEE models as: 
\begin{equation*}
\bar{\bbeta} = \dfrac{1}{M} \sum_{m=1}^M \hat{\bbeta}^m
\end{equation*}

Afterwards, longitudinal prediction on disease outcomes for a new subject $k$ at time $t$ in the test set is made using:

\begin{equation} \label{eq:eq2}
\hat{y}_{kt} = g^{-1} \left(
\hat{\mu} + \tilde{\bX}_{kt} \bar{\balpha} + \bz_k \bar{\bgamma}
\right)
\end{equation}

\medskip

\noindent
where $\tilde{\bX}_{kt} = \bX_{kt} \hat{\bS}$ represents the selected features from subject $k$ at time $t$ and the selection is dependent only on the training set (as $\hat{\bS}$ is estimated from the training data); $\bar{\balpha}$ is of the same dimension with the original number of smartphone features and it has $q$ non-zero entries in locations corresponding to the $q$ selected smartphone features in $\bar{\bbeta}$ and zeros in the remaining entries; similarly, $\bar{\bgamma}$ is of the same dimension of total covariates and it has $p-q$ non-zero entries corresponding to the $p-q$ selected covariates in $\bar{\bbeta}$ and zeros in the remaining entries.

It is worthwhile mentioning that, besides averaging $M$ sets of parameters to yield a unified predictive model, one could alternatively average multiple sets of predictions from each of the $M$ models. These two types of averaging techniques, however, yielded very similar outcomes in this study (see \textbf{Supplementary Information}), and we proceed with the predictive model described in \textbf{Eq.} \eqref{eq:eq2}. 

The data organization, imputation, model development, and parameter ensemble are summarized in \textbf{Fig.} \ref{Fig_1}. As a remark, we did not perform MI for the test data to ensure that (1) the evaluation of model performance was only done on the observed (test) data; and (2) the ability of the model to conduct personalized disease assessment on novel PwMS was independent of the imputation mechanism.

\subsection{Personalized longitudinal disease assessment and its improvement} \label{sec:2.6}

Longitudinal disease assessment needs to address two important problems: extrapolation and personalization. For extrapolation, it needs to identify behavioral markers that are present in a broad patient population so that they can be extended to novel samples. For personalization, it needs to account for idiosyncratic information that is unique to each patient and is therefore potentially not included in the population-level parameters.

In methods development, these two sets of problems translate to the following tasks. 

\begin{figure}[h]
\includegraphics[width=90mm]{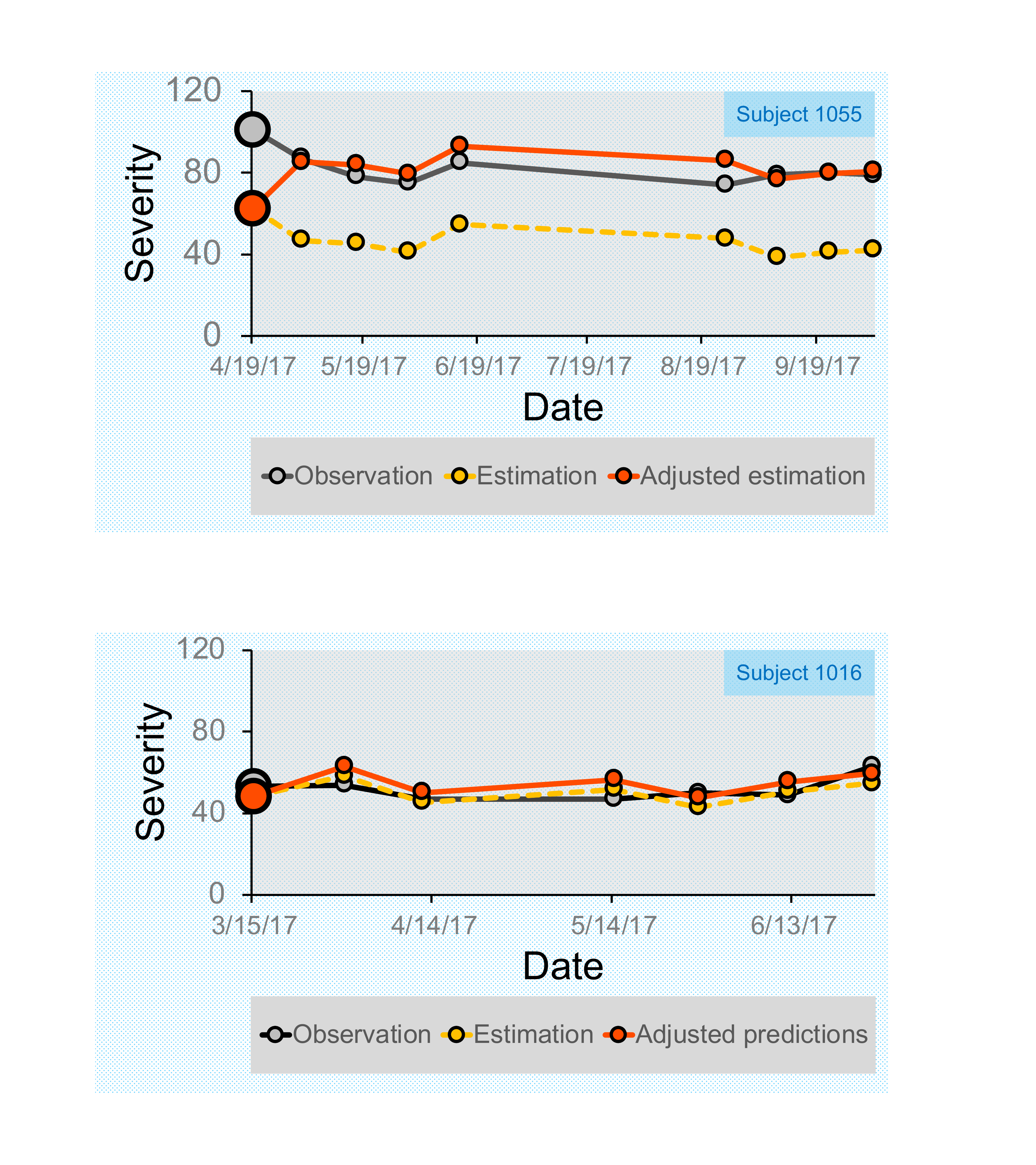}
\smallskip
\noindent\colorbox{olivergray}{
\parbox[c]{8.3cm} {Two individuals, one with severe MSIS-29 scores (top panel) and another with minor scores (bottom panel), are shown. The grey circles represent observed longitudinal scores. The yellow circles are estimated scores using the ensemble generalized estimation equations (GEE) parameters. The red circles are estimated scores further fine-tuned using data from the first day (see text for details). 
}
}
\caption{Personalized longitudinal disease assessment using smartphones.}
\label{Fig_3}
\end{figure}

\textbf{Task 1}: Can we discover features whose patterns over time are associated with longitudinal outcomes and such association is generally present in all PwMS in the training sample? If so, they may be useful to make forecasts for novel individuals (see \textbf{Fig.} \ref{Fig_3}); this can be further verified using out-of-sample testing. 

\textbf{Task 2}: Can we further adjust the longitudinal predictions to reflect individual differences? Stated differently, the trained parameters during GEE model development reflect the relationship between the features and the outcomes specific to the training sample; if we were to use these parameters to make forecasts about outcomes in novel PwMS, they may not fully reveal the relationship between the features and outcomes in novel individuals. For example, a population-level model in practice may underestimate disease severity in novel PwMS during out-of-sample testing (see \textbf{Figs.} 3 and 5). On the other hand, if one considers subject-specific parameters (such as that in the mixed effect models) or incorporates prior information about a subject (such as that in Bayesian learning) for the training model, such parameters or priors still only reflected individual information with regards to the training sample. Naturally, one would want to make an individualized adjustment to the population-level model (whose parameters were developed on the training sample) to incorporate subject-specific information that is contained in the test sample. But constructing an (independent) subject-specific model for the test sample to include idiosyncratic information may over-fit the test data.

To consider and balance extrapolation and personalization, we added the following subject-level fine-tuning into the longitudinal model:

\begin{equation} \label{eq:eq3}
\hat{y}_{kt}^{\text{adj}} = 
\begin{cases}
      \hat{y}_{kt} & t=1 \\
      \hat{y}_{kt} + ( y_{k1} - \hat{y}_{k1} ) & t \geq 2 
    \end{cases}
\end{equation}

\medskip

\noindent
where $\hat{y}_{kt}$ is the estimated outcome for a new subject $k$ at time $t$ using \textbf{Eq.} \eqref{eq:eq2}, $y_{k1}$ is the earliest ($t=1$) observed outcome for subject $k$ and $\hat{y}_{k1}$ is its estimation, and $\hat{y}_{kt}^{\text{adj}}$ is the fine-tuned estimate.

In words, the above correction means that after disease estimation has been made using the population-level parameters from the ensemble GEE on novel PwMS (see dashed lines in \textbf{Fig.} \ref{Fig_3}), we make a subject-level correction (by adjusting a constant $(y_{k1} - \hat{y}_{k1})$ on estimates made from day two and beyond using disease information from the first day (see solid red lines in \textbf{Fig.} \ref{Fig_3}).

There are a few reasons for using one day’s data for fine-tuning (instead of using multiple days) in this study. First, when there are only a small number of data points recorded, using data from multiple days for tuning would result in fewer days of data left for testing (and could potentially artificially boost test results). Second, our analyses demonstrated that fine-tuning using data from one day indeed led to improved personalized disease monitoring while avoiding over-fitting the test data. The reason that the fine-tuning avoided over-fitting was twofold. (a) It did not use any information from the outcomes obtained after day one. (b) It only adjusted a constant $(y_{k1} - \hat{y}_{k1})$ on the prediction. Had the estimated longitudinal trend been inaccurate using \textbf{Eq.} \eqref{eq:eq2} (see the yellow line in \textbf{Fig.} \ref{Fig_3}), a constant adjustment in \textbf{Eq.} \eqref{eq:eq3} would not improve the trend prediction. Finally, when longer test data become available for each individual in the future, one can potentially further improve the personalization by using a fine(r)-tuning based on more data or by extracting a subject-level prior for each novel subject using data obtained during an initial period. But our message of considering and balancing extrapolation and fine-tuning from the current proof-of-concept study remains.

\subsection{Optimal number of imputed datasets} \label{sec:2.7}

When employing MI, an interesting question arises. How many sets of imputed data are needed? Although the percentage of missing data in our MS data set is relatively small ($\gamma=1.42\%$), missing just one entry of an important feature on a particular day would make it impossible to make a disease assessment on that day. More concretely, this question needs to be addressed in two aspects. First, how many sets of imputations would yield high prediction accuracy – this needs to be evaluated empirically. Second, theoretically, how can we determine that the so-called ``high prediction accuracy'' is efficiently high?

To examine the optimal number of sets of imputations empirically, we selected a variety number of choices for $M$ and calculated their out-of-sample prediction accuracies (quantified as the correlation between true and estimated disease severity scores) (see \textbf{Fig.} \ref{Fig_2}). Our results showed that the prediction accuracy with MI was generally higher than it with missing values. Additionally, once more than $15$ imputations were made, the accuracy became relatively stable. 

Next, the number of imputations and efficiency have the following approximate relationship \cite{rubin2004multiple}: 

\begin{equation} \label{eq:eq4}
\text{Efficiency} =(1 + \dfrac{\gamma}{M})^{-1}   
\end{equation}

\medskip

\noindent
where $\gamma$ refers to the percentage of missing data, and $M$ denotes the number of imputations. When setting $M=15$, the theoretical efficiency was $99.9\%$ using the above formula. Since adding more imputations had not significantly improved prediction accuracy or efficiency and making too many sets of imputations would incur unnecessary computing time, we chose $M=15$ for the remaining of the paper.

\section{Results}

In this section, we present longitudinal MS prediction results using smartphone data. Throughout, to deal with uncertainty resulting from imputation, we generated $15$ imputed values for each missing value. That is, for each experiment, $15$ sets of complete data containing imputed values were used to train the model. Fifteen GEE models were subsequently fit for each of the imputed, and now full, training data set. The resulting $15$ sets of estimated parameters were pooled to form a unified, ensemble GEE predictive model. The final predictive model was then used to forecast disease outcomes in independent test PwMS. To ensure that the out-of-sample prediction was not affected by imputation mechanisms, we only considered imputation for the training data and evaluated the model performance on independent PwMS only on days where disease scores were available. 

\begin{table*}[]
\resizebox{\textwidth}{!}{%
\begin{tabular}{c|cccccc|cccccc|}
\cline{2-13}
\multicolumn{1}{l|}{} &
  \multicolumn{6}{c|}{{\color[HTML]{167AAB} Mean prediction}} &
  \multicolumn{6}{c|}{{\color[HTML]{167AAB} Longitudinal prediction}} \\ \cline{2-13} 
\multicolumn{1}{l|}{} &
  \multicolumn{3}{c|}{\cellcolor[HTML]{D5FFFF}EDSS} &
  \multicolumn{3}{c|}{\cellcolor[HTML]{76D6FF}MSIS-29} &
  \multicolumn{3}{c|}{\cellcolor[HTML]{D5FFFF}EDSS} &
  \multicolumn{3}{c|}{\cellcolor[HTML]{76D6FF}MSIS-29} \\ \cline{2-13} 
\multicolumn{1}{l|}{} &
  \multicolumn{1}{c|}{\cellcolor[HTML]{FFD7D6}\begin{tabular}[c]{@{}c@{}}Mixed \\ effect\end{tabular}} &
  \multicolumn{1}{c|}{\cellcolor[HTML]{FFACA9}GLM} &
  \multicolumn{1}{c|}{\cellcolor[HTML]{FF7E79}GEE} &
  \multicolumn{1}{c|}{\cellcolor[HTML]{FFD7D6}\begin{tabular}[c]{@{}c@{}}Mixed \\ effect\end{tabular}} &
  \multicolumn{1}{c|}{\cellcolor[HTML]{FFACA9}GLM} &
  \cellcolor[HTML]{FF7E79}GEE &
  \multicolumn{1}{c|}{\cellcolor[HTML]{FFD7D6}\begin{tabular}[c]{@{}c@{}}Mixed\\ effect\end{tabular}} &
  \multicolumn{1}{c|}{\cellcolor[HTML]{FFACA9}GLM} &
  \multicolumn{1}{c|}{\cellcolor[HTML]{FF7E79}GEE} &
  \multicolumn{1}{c|}{\cellcolor[HTML]{FFD7D6}\begin{tabular}[c]{@{}c@{}}Mixed \\ effect\end{tabular}} &
  \multicolumn{1}{c|}{\cellcolor[HTML]{FFACA9}GLM} &
  \cellcolor[HTML]{FF7E79}GEE \\ \hline
\rowcolor[HTML]{F7EED9} 
\multicolumn{1}{|c|}{\cellcolor[HTML]{F2F2F2}{\color[HTML]{404040} \textit{r}}} &
  \multicolumn{1}{c|}{\cellcolor[HTML]{F7EED9}0.414} &
  \multicolumn{1}{c|}{\cellcolor[HTML]{F7EED9}0.486} &
  \multicolumn{1}{c|}{\cellcolor[HTML]{F7EED9}0.493} &
  \multicolumn{1}{c|}{\cellcolor[HTML]{F7EED9}0.502} &
  \multicolumn{1}{c|}{\cellcolor[HTML]{F7EED9}0.639} &
  0.678 &
  \multicolumn{1}{c|}{\cellcolor[HTML]{F7EED9}0.417} &
  \multicolumn{1}{c|}{\cellcolor[HTML]{F7EED9}0.475} &
  \multicolumn{1}{c|}{\cellcolor[HTML]{F7EED9}0.542} &
  \multicolumn{1}{c|}{\cellcolor[HTML]{F7EED9}0.628} &
  \multicolumn{1}{c|}{\cellcolor[HTML]{F7EED9}0.647} &
  0.682 \\ \hline
\rowcolor[HTML]{F3B46C} 
\multicolumn{1}{|c|}{\cellcolor[HTML]{BFBFBF}{\color[HTML]{404040} \textit{r} (adjusted)}} &
  \multicolumn{1}{c|}{\cellcolor[HTML]{F3B46C}*} &
  \multicolumn{1}{c|}{\cellcolor[HTML]{F3B46C}*} &
  \multicolumn{1}{c|}{\cellcolor[HTML]{F3B46C}*} &
  \multicolumn{1}{c|}{\cellcolor[HTML]{F3B46C}0.837} &
  \multicolumn{1}{c|}{\cellcolor[HTML]{F3B46C}0.878} &
  0.880 &
  \multicolumn{1}{c|}{\cellcolor[HTML]{F3B46C}*} &
  \multicolumn{1}{c|}{\cellcolor[HTML]{F3B46C}*} &
  \multicolumn{1}{c|}{\cellcolor[HTML]{F3B46C}*} &
  \multicolumn{1}{c|}{\cellcolor[HTML]{F3B46C}0.792} &
  \multicolumn{1}{c|}{\cellcolor[HTML]{F3B46C}0.798} &
  0.804 \\ \hline
\rowcolor[HTML]{F7EED9} 
\multicolumn{1}{|c|}{\cellcolor[HTML]{F2F2F2}{\color[HTML]{404040} MSE}} &
  \multicolumn{1}{c|}{\cellcolor[HTML]{F7EED9}1.41} &
  \multicolumn{1}{c|}{\cellcolor[HTML]{F7EED9}1.203} &
  \multicolumn{1}{c|}{\cellcolor[HTML]{F7EED9}1.186} &
  \multicolumn{1}{c|}{\cellcolor[HTML]{F7EED9}25.668} &
  \multicolumn{1}{c|}{\cellcolor[HTML]{F7EED9}22.631} &
  22.307 &
  \multicolumn{1}{c|}{\cellcolor[HTML]{F7EED9}1.416} &
  \multicolumn{1}{c|}{\cellcolor[HTML]{F7EED9}1.142} &
  \multicolumn{1}{c|}{\cellcolor[HTML]{F7EED9}1.091} &
  \multicolumn{1}{c|}{\cellcolor[HTML]{F7EED9}21.546} &
  \multicolumn{1}{c|}{\cellcolor[HTML]{F7EED9}19.795} &
  19.527 \\ \hline
\rowcolor[HTML]{F3B46C} 
\multicolumn{1}{|c|}{\cellcolor[HTML]{BFBFBF}{\color[HTML]{404040} MSE (adjusted)}} &
  \multicolumn{1}{c|}{\cellcolor[HTML]{F3B46C}*} &
  \multicolumn{1}{c|}{\cellcolor[HTML]{F3B46C}*} &
  \multicolumn{1}{c|}{\cellcolor[HTML]{F3B46C}*} &
  \multicolumn{1}{c|}{\cellcolor[HTML]{F3B46C}15.611} &
  \multicolumn{1}{c|}{\cellcolor[HTML]{F3B46C}13.197} &
  13.011 &
  \multicolumn{1}{c|}{\cellcolor[HTML]{F3B46C}*} &
  \multicolumn{1}{c|}{\cellcolor[HTML]{F3B46C}*} &
  \multicolumn{1}{c|}{\cellcolor[HTML]{F3B46C}*} &
  \multicolumn{1}{c|}{\cellcolor[HTML]{F3B46C}14.833} &
  \multicolumn{1}{c|}{\cellcolor[HTML]{F3B46C}14.961} &
  14.785 \\ \hline
\end{tabular}%
}

\smallskip

\noindent
\justifying{*Most people with MS (PwMS) only have one to two EDSS scores. It is therefore impractical to make further adjustments using data from day one.}

\smallskip

\noindent
\noindent\colorbox{olivergray}{
\parbox[c]{17.8cm} {Left: Model comparison for mean outcome prediction. GEE model is compared with two baseline models, a mixed-effect model and a GLM, to predict both average EDSS and MSIS-29 scores. The prediction accuracy is evaluated using two metrics: the correlation (or \textit{r}) and the mean squared errors (MSE). Right: Model comparison for longitudinal outcome prediction. The same analysis is performed for longitudinal disease prediction, where the disease outcomes are longitudinally observed individual EDSS and MSIS-29 scores.}
}
\smallskip

\caption{Model comparison.}
\label{Table1}
\end{table*}

To demonstrate the efficacy of the proposed framework, we compared it with mixed-effect and GLM models in the same modeling and test strategy. For each model, we applied it to predict both the averaged outcomes and longitudinal outcomes, and for both EDSS and MSIS-29 scores. We recorded the accuracy statistics (correlation and MSE) from each model (see \textbf{Table} \ref{Table1}). Our results showed that for both disease types, GEE outperformed the mixed-effect model and was slightly better than GLM in mean assessment and longitudinal assessment. Since there were far fewer longitudinal measurements of EDSS score per individual compared to those of MSIS-29 score and the main theme herein is on longitudinal disease severity prediction, in the following we will focus on examining the performance of the proposed model with regards to longitudinal MSIS-29 score prediction.

\begin{figure}[h]
\includegraphics[width=90mm]{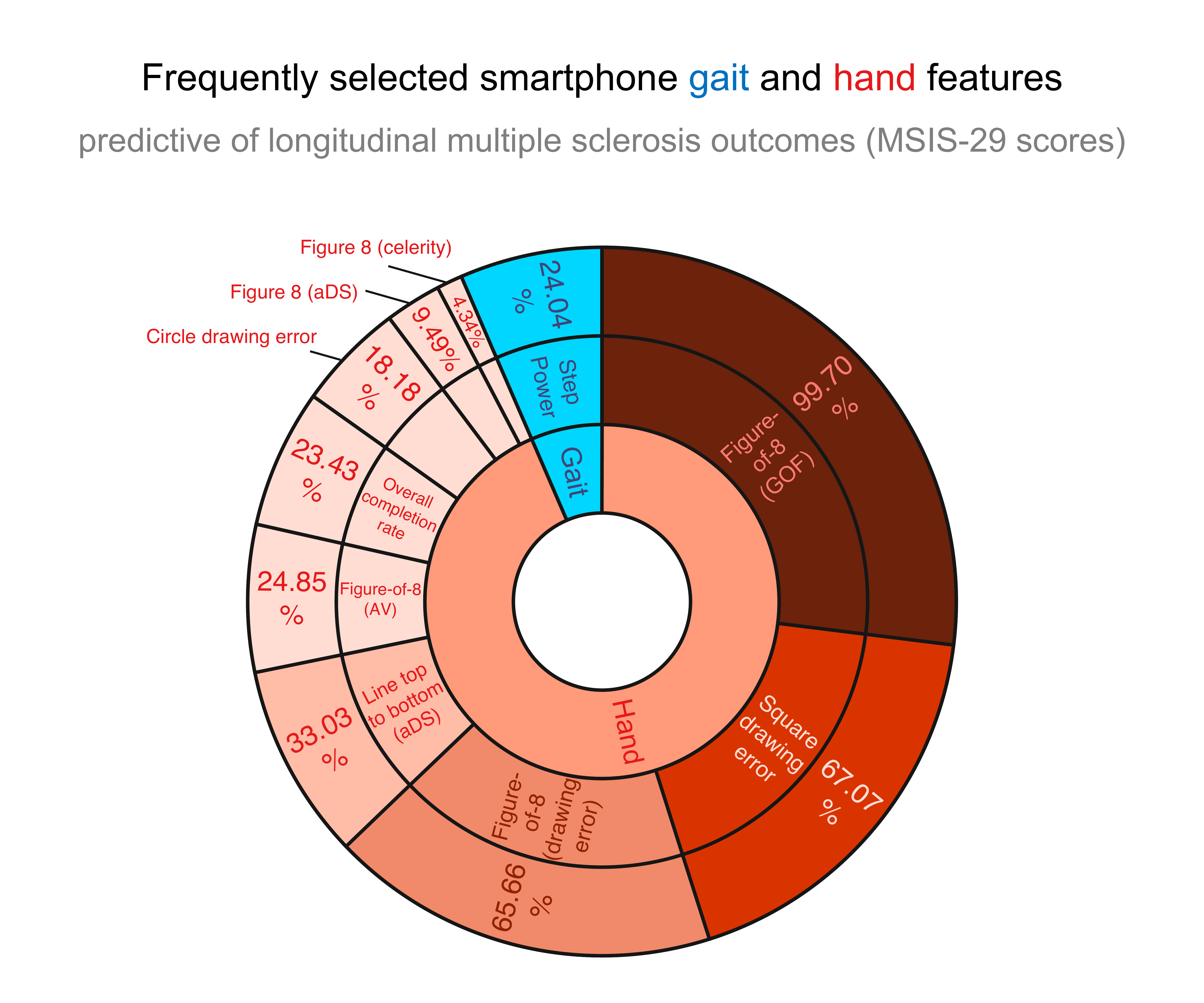}
\smallskip
\noindent\colorbox{olivergray}{
\parbox[c]{8.3cm} {The distribution of selected gait and hand features. One thousand bootstrap experiments were performed to discover which gait and hand features were consistently selected for longitudinal MSIS-29 prediction. The selected gait and hand features were plotted on a pie chart: from inner circle to outer circle were feature category (gait \textit{vs.} hand features), feature name, and the frequency each feature was selected from one thousand bootstrap experiments. Gait feature(s) were coded in blue and hand feature(s) were coded in red (see text for details on individual features). Note although six features were allowed during each bootstrap, the selected (six) features may vary across different bootstraps; but noticeably some features were more frequently selected.
}
}
\caption{Key features in smartphone-based remote longitudinal outcome prediction for multiple sclerosis patients.}
\label{Fig_4}
\end{figure}

Our results suggest features related to gait and upper extremity functions are useful to assess MS longitudinally (see \textbf{Fig.} \ref{Fig_4}). Specifically, the following DaS features \cite{creagh2020asmartphone} are consistently selected: ``Figure-of-8 GOF'', which is the root-mean-squared error obtained from a linear regression between the reference trace of the figure-of-8 and the drawn trace (\textit{i.e.}, goodness of fit [GOF]); ``Figure-of-8 AV'', which is the angular drawing velocity (AV); ``Figure-of-8 trace celerity'', which is the ratio of trace accuracy over drawing time; ``Figure-of-8 drawing error'', ``Circle drawing error'', and ``Square drawing error'', which quantify the error made when drawing a figure-of-8, circle, and square, respectively; ``Figure-of-8 aDS'' and ``Line top to bottom aDS'', which quantifies the absolute drawing speed (aDS) when drawing the figure-of-8 and line top to bottom, respectively; ``Overall completion rate'', which describes the proportion of successfully drawn shapes. From the 2MWT \cite{creagh2020bsmartphone}, ``Step power'' is consistently selected. It quantifies the power, or energy, invested per step during a 2MWT. 

\begin{figure*}[h]
\includegraphics[width=180mm]{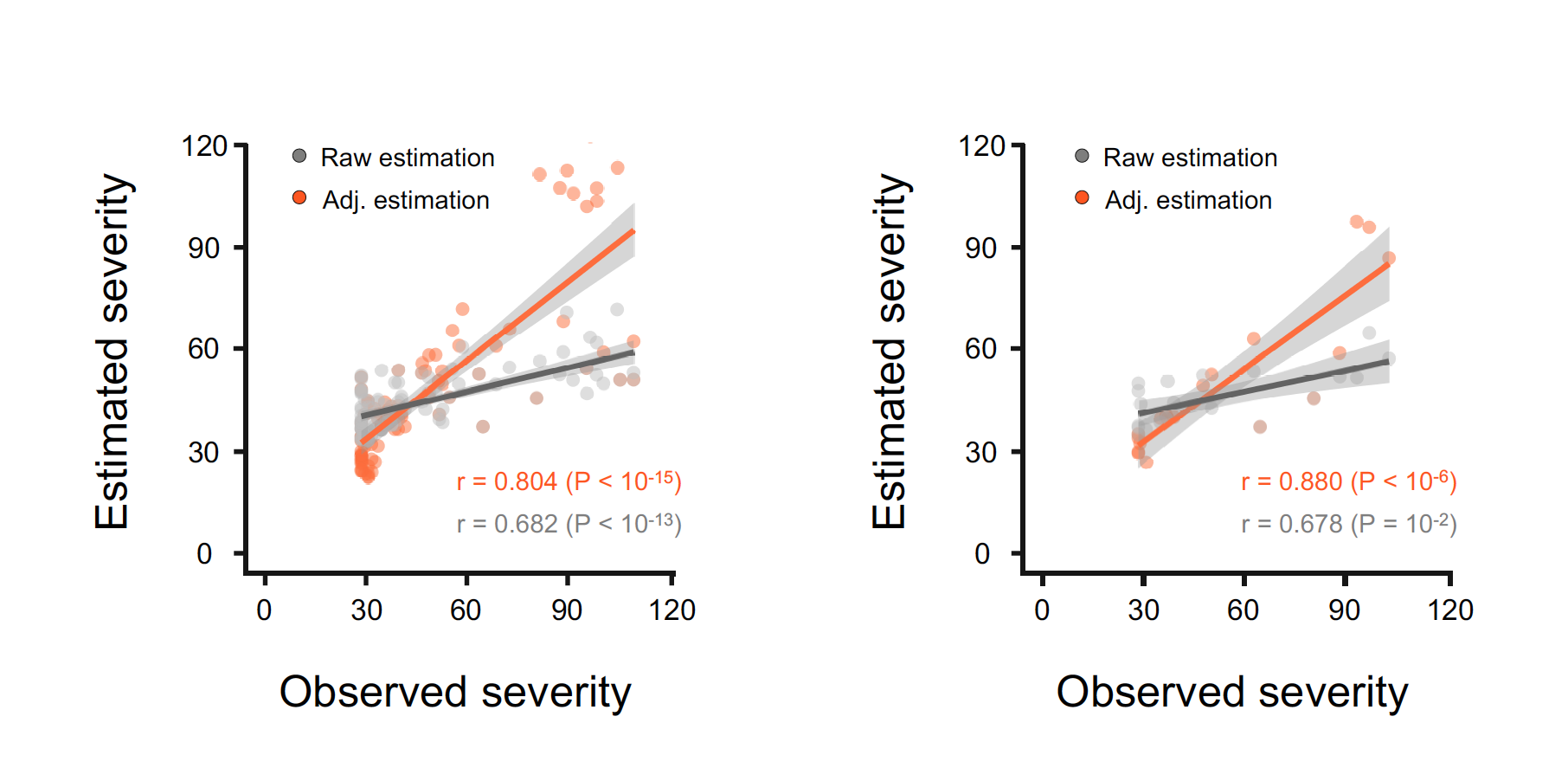}
\smallskip
\noindent\colorbox{olivergray}{
\parbox[c]{17.8cm} {Left: The estimated against observed MSIS-29 scores for previously unseen people with MS (PwMS). The grey dots correspond to the (raw) estimations using ensemble GEE in \textbf{Eq.} \eqref{eq:eq2}. The orange dots are the fine-tuned estimations using \textbf{Eq.} \eqref{eq:eq3}. Right: To examine whether the significant correlations observed in the left figure are due to repeated measurements per subject, the individual mean estimations are plotted against individual mean observations. The orange and grey lines represent the linear association between estimated and observed values.
}
}
\caption{Estimating multiple sclerosis (MS) severity over time in novel individuals.}
\label{Fig_5}
\end{figure*}

Overall, the estimated longitudinal MSIS-29 scores in the novel test sample were significantly correlated with their observed counterparts ($r = 0.80$, $p < 0.001$) (see left of \textbf{Fig.} \ref{Fig_5}). It, however, remained possible that the significant association was boosted by having repeated measurements in the test set. To check for this possibility, we took the means of the predicted and observed outcomes for each individual (thus there was only one estimated mean score and one predicted mean score for each participant) and calculated the association again; the result remained significant ($r = 0.88, p < 0.001$).

\subsection{Bootstrap experiments on evaluating predictions at the group level} \label{sec:3.1}

Since training and test samples were obtained from a random split of the data, predictions made on some splits may perform better than those made on others. To evaluate the general performance of our model, we performed $1,000$ bootstrap experiments. Specifically, each bootstrap began with a random draw of the data. The PwMS in the sampled data were subsequently split into $70\%$ (training) and $30\%$ (test), where repeated measurements of a subject were either in the training set or the test set. Next, for each bootstrap sample, multiple imputation ($M=15$) was made on its training set, followed by model development using the GEE on each of the imputed full set. Subsequently, parameters from the $M$ models were pooled to form an ensemble GEE model that was specific to each bootstrap training sample, with model performance evaluated on the test data. Each bootstrap experiment ended with one correlation value between the estimated and observed outcomes regarding the test set. Together, we obtained $1,000$ correlations values from all bootstrap experiments. In parallel, we performed another $1,000$ bootstrap experiments but with the personalized fine-tuning included as outlined in \textbf{Eq.} \eqref{eq:eq3}.

\begin{figure}[h]
\includegraphics[width=90mm]{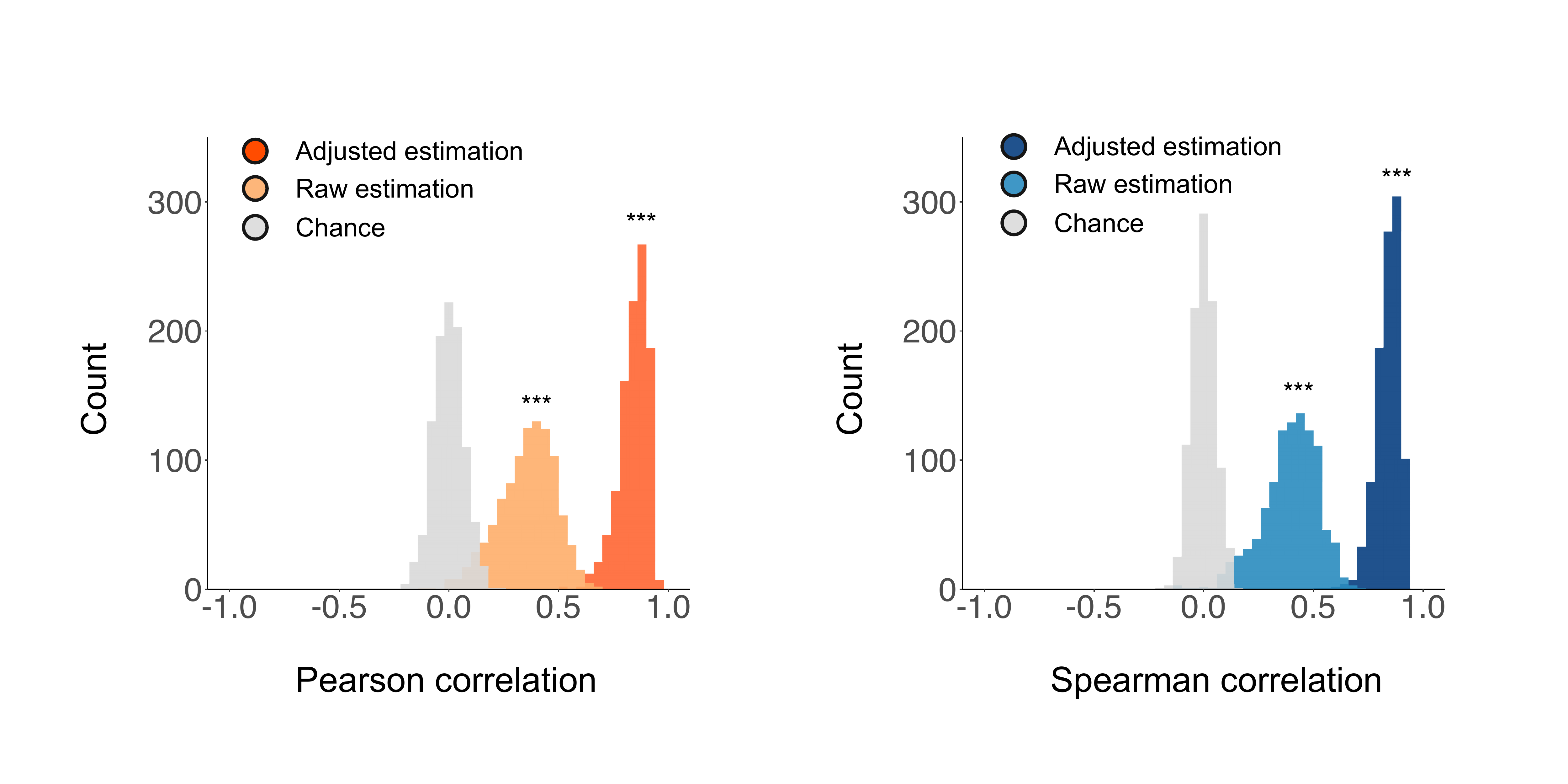}
\smallskip
\noindent\colorbox{olivergray}{
\parbox[c]{8.3cm} {During each bootstrap test, data from people with MS (PwMS) were randomly drawn from the sample. Each selected sample was then randomly split into a training set (containing $70\%$ of the total samples) and a test set (containing $30\%$ of the total samples). Next, MI was performed, and an ensemble GEE was learned from the training data. The trained model was then evaluated on the test data. Subsequently, the (longitudinal) correlation between the estimated and observed disease scores was calculated. This was repeated $1,000$ times, producing $1,000$ correlations. The light orange and light blue shaded histograms represent the distributions of Pearson and Spearman correlations from $1,000$ bootstrap tests; the dark orange and dark blue shaded histograms represent the distributions of the fine-tuned personalized prediction outcomes. For comparison, two null distributions (the grey shaded histograms) were generated from a normal distribution with zero mean and the same standard deviations as the bootstrap Pearson and Spearman correlations.
} 
}
\caption{Bootstrap results of out-of-sample testing performance for MSIS-29 score prediction.}
\label{Fig_6}
\end{figure}

The bootstrap results showed that, first, the proposed method using GEE and MI was reproducible across all bootstrap samples with reasonable overall disease assessment results (with an average correlation of $0.34$), and second, the adjusted personalized disease assessment further improved estimation accuracy (with an average correlation of $r = 0.81$) (see \textbf{Fig.} \ref{Fig_6}).

Yet, one may still question, how can one be certain that the above analyses justify the efficacy of the model for longitudinal prediction? It remained possible that, if the dataset contained many individuals without much longitudinal variability in their (within-subject) scores, then the above analyses only showed that the model was useful in estimating the average scores. To further examine the model performance in longitudinal disease prediction, we performed two additional analyses detailed in the next sub-section.

\subsection{Evaluating the longitudinal model performance regarding predicting the MSIS-29 scores} \label{sec:3.2}

\begin{figure}[h]
\includegraphics[width=90mm]{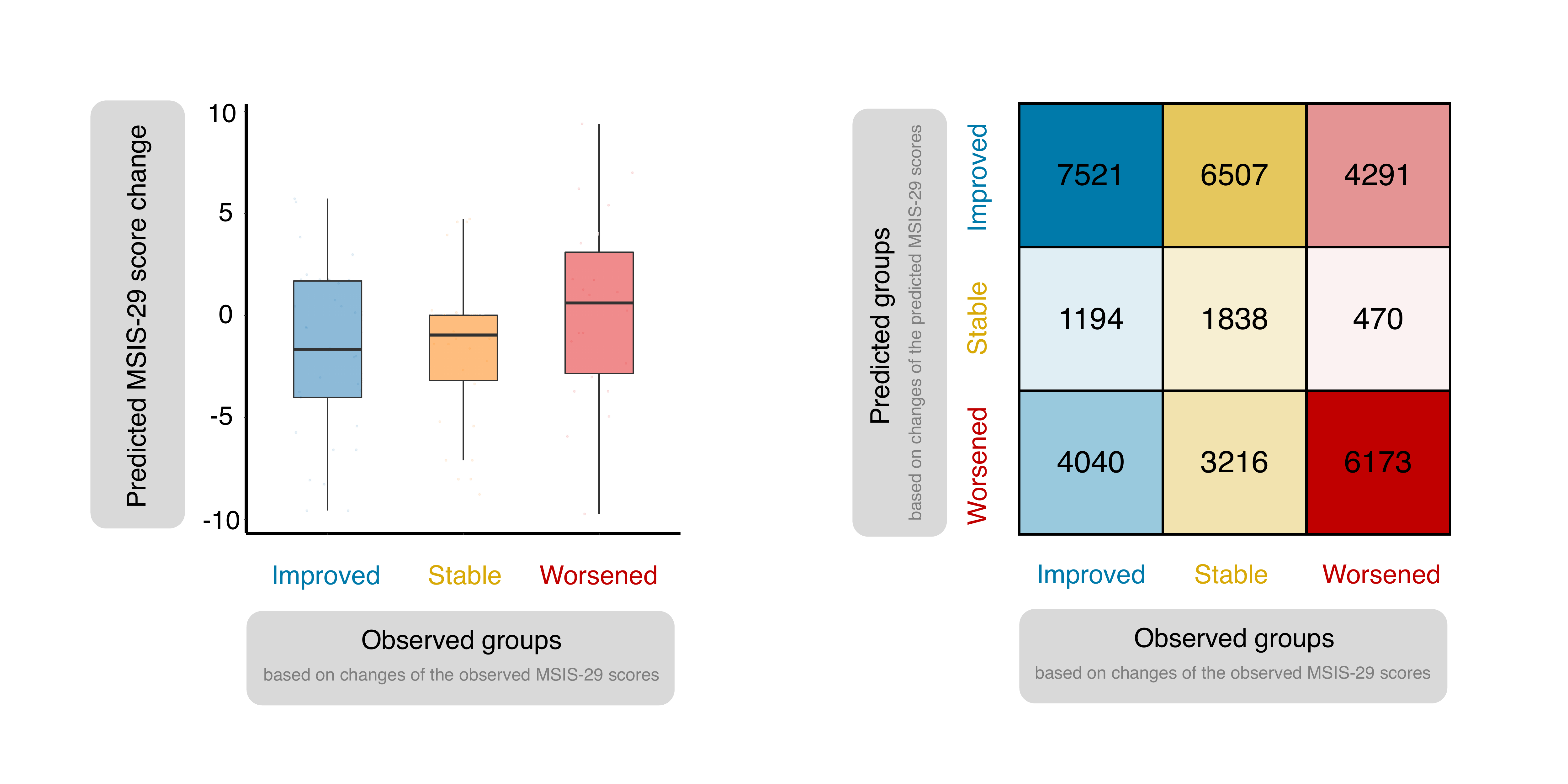}
\smallskip
\noindent\colorbox{olivergray}{
\parbox[c]{8.3cm} {Left: We performed a leave-one-subject-out cross-validation (LOOCV) analysis and plotted the predicted disease score changes against patients' disease categories. Right: We performed a bootstrap analysis and presented the number of patients falling into each predicted \textit{vs.} observed group (see text for details). 
}
}
\caption{Results of predicted score change against observed groups in out-of-sample people with MS (PwMS).}
\label{Fig_7}
\end{figure}

We first considered an additional analysis using leave-one-subject-out cross-validation (LOOCV) to determine whether the predicted disease scores captured longitudinal trends. Let's begin by denoting $\tau$ ($\tau>0$) as the grouping threshold. First, we group the PwMS into three categories: (1) Improved, (2) Stable, and (3) Worsened. The categories were determined based on whether the change in MSIS-29 scores at the end of the study from baseline was smaller than $-\tau$ units (denoted as $\Delta<-\tau$, indicating improvement), within plus/minus $\tau$ units (denoted as $\Delta \in [-\tau, \tau]$, indicating stable), or greater than $+\tau$ units (denoted as $\Delta > + \tau$, indicating worsening). We set threshold $\tau = 1/2$ for demonstration purposes. If the directions of estimated score changes agreed with the observed categories, then it suggests that the model was able to pick up the longitudinal disease trend. Although the means of the three groups were not pairwise significant (this is in part due to the small sample size, and in part due to the small longitudinal changes during a period of six months), the LOOCV results suggested that the predicted score changes were generally in line with the observed categories (see \textbf{Fig.} \ref{Fig_7}). Note that a larger $\tau$ would yield more observations to be classified into the stable group; but across different sets of grouping thresholds (\textit{i.e.}, $\tau=1$, $2$, and $3$), we observed similar results.

We then performed an additional $1,000$ bootstrap experiments (with the same setting in \textbf{Section} \ref{sec:3.1}) as follows. First, individuals were classified into improved, stable, or worsened groups based on whether the disease scores at the end of the study were at least a half unit smaller than, similar to (within plus/minus a half unit), or at least a half unit larger than, the score at the baseline. Longitudinal disease prediction was carried out and the difference between the predicted baseline and the scores at the end of the study for each individual was calculated. We then counted the numbers of individuals falling in each predicted category against those falling in each observed category (see \textbf{Fig.} \ref{Fig_7}). Note that some patients showed improved scores due to therapeutic effects or recovery from relapse events.

We next sought to quantify the longitudinal prediction performance. To that end, we calculated the repeated measures correlation ($rmcorr$) \cite{bakdash2017repeated} for the outcome-of-sample PwMS to estimate the within-individual association between longitudinal observations and their estimates for multiple individuals. The $rmcorr$ was calculated using the publicly available $R$ package $rmcorr$ \cite{bakdash2017repeated}. Specifically, we considered the $rmcorr$ for estimated and observed baseline and the MSIS-29 scores at the end of the study: a positive $rmcorr$ would suggest that there was potential evidence that the model was able to detect the longitudinal within-individual disease score change between the baseline and at the end of the study. To account for variability due to splitting mechanisms (between training and test samples), we performed $1,000$ bootstrap experiments and obtained $1,000$ out-of-sample $rmcorr$ values (mean $rmcorr=0.14$, $st.d.=0.19$) and Pearson correlations (mean $r=0.42$, $st.d.=0.16$), respectively. We note that the variability in the $rmcorr$ estimate does not allow us to rule out the possibility of there being no positive longitudinal correlation; the estimate of $P(rmcorr<0)$ from the bootstrap samples, however, was only $0.22$. Taken together, these results suggested that the model was overall promising to capture the longitudinal within-individual disease trend, with limitations discussed below.

\section{Discussion}
The results from \textbf{Sections} \ref{sec:3.1} and \ref{sec:3.2} suggest that the usefulness of the proposed model is twofold. First, it balances group- and subject-level information. The parameters were learned from the training sample and contained information that may be manifested at the group level and can be extrapolated to previously unseen PwMS (see the light orange histogram in \textbf{Fig.} \ref{Fig_6}). The novel PwMS, however, may contain information that may not be captured by the population parameters. The subject-specific fine-tuning using data from day one may narrow the gap (see the dark orange histogram in \textbf{Fig.} \ref{Fig_6}). Nevertheless, if the PwMS have scores that do not vary much over time, the model may underperform the one using the data from day one alone. Said differently, if the scores of PwMS do not vary much longitudinally, one may as well use a participant's score from day one, say, $50$, to predict the future scores; that is, a string of constant scores. Yet, the analyses in \textbf{Section} \ref{sec:3.2} suggest that even when the individual scores had not progressed significantly, the model seemed to be able to pick up the longitudinal trend. Previous findings that it is possible to obtain longitudinal trend approximate to EDSS of multiple sclerosis using gait data (2MWT) collected by smartphones \cite{creagh2022longitudinal}. Our work confirms this, and further expands to the longitudinal proxies to both EDSS and MSIS-29 scores using gait (2MWT) and upper extremity function (DoS) features. Such converging evidence suggests an additional attractive (longitudinal) property of our model; it also suggests that when there are even minor disease dynamics, one should consider a longitudinal model and should refrain from making extrapolations that, since the disease (for an individual) is relatively stable, one can rely on the first disease score to infer (a string of constant) future scores. Finally, when dealing with longitudinal prediction with different levels of missing data, \textbf{Eq.} \ref{eq:eq4} provides a helpful reference to which one can choose the number of imputations based on the amount of missing data to achieve a desirable efficiency.

There are a few limitations of our study. First, there is a limitation in current data, as they were acquired during a relatively short period wherein the longitudinal scores of PwMS might not significantly change. Future studies can further verify the proposed method in data collected over a much longer period, during which more changes in disease activity can be expected. Additionally, this study consists of subjects with relatively mild levels of MS-related functional impairment. Future studies may further consider a broader population including healthy controls, patients with mild MS symptoms, and patients with severe, potentially fluctuating disease profiles. Beginnings are already being made; see, for example, \cite{van2019floodlight, roy2022disability}. Further, based on previous findings and neurobiological insights, in this study we were mainly interested in examining the efficacy of gait and upper extremity functions in predicting longitudinal MS progress. Although our findings have suggested the utility of these features in longitudinal MS prediction, we have naturally left out a large territory where other feature modalities, such as dexterity tests (\textit{e.g.}, pinching test), cognitive test data (\textit{e.g.}, the Symbol Digit Modalities Test (SDMT)), U-turn test, and the passively collected digital data, may also be useful to assess MS over time. Further research may, on the one hand, explore the longitudinal prediction performance bestowed by each feature modality, and, on the other hand, investigate whether, and if so, to what extent, one may improve longitudinal MS prediction by integrating multivariate multi-modal features. Finally, although our model was able to significantly predict subject-specific mean scores and modestly identify the subject-specific longitudinal disease trend, the results of the latter were not statistically significant. This may be due, in part, to the short study period and small sample size and, in part, to the limitation of the model. Future studies should verify the proposed approach on larger datasets and explore additional modeling approaches to improve the proposed framework.

\section{Conclusions}
Personalized longitudinal assessment of MS disease has the potential to inform clinical decisions, and thereby improve treatment outcomes. Smartphone sensor-based assessments offer a new cost-efficient approach to remotely and frequently assessing MS-related functional ability that can complement standard clinical assessments \cite{montalban2021smartphone}. Smartphone devices are widely available and generate, through their embedded sensors, highly granular and meaningful data suitable for longitudinal modeling of MS. Before such sensor-based assessment can be routinely deployed in clinical practice, it is important to evaluate whether, and if so, to what extent, they can distinguish between-individual differences in disease profiles and uncover within-individual disease courses longitudinally. 

In this study, we developed an automated personalized longitudinal framework to assess MS over time. The framework combines MI, GEE, ensemble learning, and subject-specific fine-tuning. MI was used to impute missing data entries; the ensemble GEE was employed for model development and longitudinal prediction of MS disease scores; the fine-tuning was introduced to adjust for idiosyncratic disease trajectory. 

Utilizing smartphone and clinical data obtained from PwMS, the framework showed promise to estimate individual longitudinal MS disease profiles in previously unseen PwMS. Particularly, the detected disease changes between baseline and the end of the study agreed in general with the observed changes in MSIS-29 scores. 

Taken together, our analyses proved the concept of smartphone-based, personalized MS assessment and demonstrated the potential of the proposed model in longitudinal MS evaluation. Future research needs to test the model using independent datasets and verify if the framework can be extended to evaluate other MS-related clinical outcomes. Future studies may also examine the utility of the method in MS prognosis (namely, predicting the disease before its onset) and explore whether this approach is useful to investigate and forecast other neurodegenerative diseases longitudinally.

\section{Supplementary Information}

A comparison between two types of averaging techniques for dealing with parameters estimated from multiple sets of data using multiple imputation (MI) (see \textbf{Table} \ref{Table2}).

\begin{table*}[h!]
\resizebox{\textwidth}{!}{%
\begin{tabular}{c|cccc|cccc|}
\cline{2-9}
\multicolumn{1}{l|}{} &
  \multicolumn{4}{c|}{{\color[HTML]{167AAB} Mean prediction}} &
  \multicolumn{4}{c|}{{\color[HTML]{167AAB} Longitudinal prediction}} \\ \cline{2-9} 
\multicolumn{1}{l|}{} &
  \multicolumn{2}{c|}{\cellcolor[HTML]{D5FFFF}EDSS} &
  \multicolumn{2}{c|}{\cellcolor[HTML]{76D6FF}MSIS-29} &
  \multicolumn{2}{c|}{\cellcolor[HTML]{D5FFFF}EDSS} &
  \multicolumn{2}{c|}{\cellcolor[HTML]{76D6FF}MSIS-29} \\ \cline{2-9} 
\multicolumn{1}{l|}{} &
  \multicolumn{1}{c|}{\cellcolor[HTML]{FF7E79}\begin{tabular}[c]{@{}c@{}}GEE\\ Averaging parameters\end{tabular}} &
  \multicolumn{1}{c|}{\cellcolor[HTML]{FFD7D6}\begin{tabular}[c]{@{}c@{}}GEE\\ Averaging predictions\end{tabular}} &
  \multicolumn{1}{c|}{\cellcolor[HTML]{FF7E79}\begin{tabular}[c]{@{}c@{}}GEE\\ Averaging parameters\end{tabular}} &
  \cellcolor[HTML]{FFD7D6}\begin{tabular}[c]{@{}c@{}}GEE\\ Averaging predictions\end{tabular} &
  \multicolumn{1}{c|}{\cellcolor[HTML]{FF7E79}\begin{tabular}[c]{@{}c@{}}GEE\\ Averaging parameters\end{tabular}} &
  \multicolumn{1}{c|}{\cellcolor[HTML]{FFD7D6}\begin{tabular}[c]{@{}c@{}}GEE\\ Averaging predictions\end{tabular}} &
  \multicolumn{1}{c|}{\cellcolor[HTML]{FF7E79}\begin{tabular}[c]{@{}c@{}}GEE \\ Averaging parameters\end{tabular}} &
  \cellcolor[HTML]{FFD7D6}\begin{tabular}[c]{@{}c@{}}GEE\\ Averaging predictions\end{tabular} \\ \hline
\rowcolor[HTML]{F7EED9} 
\multicolumn{1}{|c|}{\cellcolor[HTML]{F2F2F2}{\color[HTML]{404040} \textit{r}}} &
  \multicolumn{1}{c|}{\cellcolor[HTML]{F7EED9}0.493} &
  \multicolumn{1}{c|}{\cellcolor[HTML]{F7EED9}0.493} &
  \multicolumn{1}{c|}{\cellcolor[HTML]{F7EED9}0.678} &
  0.680 &
  \multicolumn{1}{c|}{\cellcolor[HTML]{F7EED9}0.542} &
  \multicolumn{1}{c|}{\cellcolor[HTML]{F7EED9}0.542} &
  \multicolumn{1}{c|}{\cellcolor[HTML]{F7EED9}0.682} &
  0.682 \\ \hline
\rowcolor[HTML]{F3B46C} 
\multicolumn{1}{|c|}{\cellcolor[HTML]{BFBFBF}{\color[HTML]{404040} \textit{r} (adjusted)}} &
  \multicolumn{1}{c|}{\cellcolor[HTML]{F3B46C}*} &
  \multicolumn{1}{c|}{\cellcolor[HTML]{F3B46C}*} &
  \multicolumn{1}{c|}{\cellcolor[HTML]{F3B46C}0.880} &
  0.880 &
  \multicolumn{1}{c|}{\cellcolor[HTML]{F3B46C}*} &
  \multicolumn{1}{c|}{\cellcolor[HTML]{F3B46C}*} &
  \multicolumn{1}{c|}{\cellcolor[HTML]{F3B46C}0.804} &
  0.803 \\ \hline
\rowcolor[HTML]{F7EED9} 
\multicolumn{1}{|c|}{\cellcolor[HTML]{F2F2F2}{\color[HTML]{404040} MSE}} &
  \multicolumn{1}{c|}{\cellcolor[HTML]{F7EED9}1.186} &
  \multicolumn{1}{c|}{\cellcolor[HTML]{F7EED9}1.186} &
  \multicolumn{1}{c|}{\cellcolor[HTML]{F7EED9}22.307} &
  22.135 &
  \multicolumn{1}{c|}{\cellcolor[HTML]{F7EED9}1.091} &
  \multicolumn{1}{c|}{\cellcolor[HTML]{F7EED9}1.091} &
  \multicolumn{1}{c|}{\cellcolor[HTML]{F7EED9}19.527} &
  19.403 \\ \hline
\rowcolor[HTML]{F3B46C} 
\multicolumn{1}{|c|}{\cellcolor[HTML]{BFBFBF}{\color[HTML]{404040} MSE (adjusted)}} &
  \multicolumn{1}{c|}{\cellcolor[HTML]{F3B46C}*} &
  \multicolumn{1}{c|}{\cellcolor[HTML]{F3B46C}*} &
  \multicolumn{1}{c|}{\cellcolor[HTML]{F3B46C}13.011} &
  12.964 &
  \multicolumn{1}{c|}{\cellcolor[HTML]{F3B46C}*} &
  \multicolumn{1}{c|}{\cellcolor[HTML]{F3B46C}*} &
  \multicolumn{1}{c|}{\cellcolor[HTML]{F3B46C}14.785} &
  14.830 \\ \hline
\end{tabular}%
}

\smallskip
\noindent
\justifying{*Most people with MS (PwMS) only have one to two EDSS scores. It is therefore impractical to make further adjustments using data from day one.}

\smallskip
\noindent
\noindent\colorbox{olivergray}{
\parbox[c]{17.8cm} {Left: Model comparison for mean outcome prediction. GEE model is compared with two baseline models, a mixed-effect model and a GLM, to predict both average EDSS and MSIS-29 scores. The prediction accuracy is evaluated using two metrics: the correlation (or \textit{r}) and the mean squared errors (MSE). Right: Model comparison for longitudinal outcome prediction. The same analysis is performed for longitudinal disease prediction, where the disease outcomes are longitudinally observed individual EDSS and MSIS-29 scores.}
}
\smallskip

\caption{Longitudinal prediction using averaged parameters versus averaged predictions. }
\label{Table2}
\end{table*}

\bibliographystyle{IEEEtran}
\bibliography{reference.bib}

\end{document}